\documentclass{article}
\usepackage[T1]{fontenc} % add special characters (e.g., umlaute)
\usepackage[utf8]{inputenc} % set utf-8 as default input encoding
\usepackage{ismir,amsmath,amsfonts,cite,url}
\usepackage{graphicx}
\usepackage{color}

\usepackage{tikz,tikz-cd,enumitem}

\newtheorem{theorem}{Theorem}[section]
\newtheorem{definition}[theorem]{Definition}

\DeclareMathOperator{\mad}{TimeError}
\DeclareMathOperator{\rmse}{TimeDev}

\DeclareMathOperator{\notemad}{NoteError}
\DeclareMathOperator{\notermse}{NoteDev}

\title{Rethinking Evaluation Methodology for \\ Audio-to-Score Alignment}

\threeauthors
  {John Thickstun} {University of Washington \\ {\tt thickstn@cs.washington.edu}}
  {Jennifer Brennan} {University of Washington \\ {\tt jrb@cs.washington.edu}}
  {Harsh Verma} {University of Washington \\ {\tt harshv@cs.washington.edu}}

\begin{document}

\maketitle
\begin{abstract}
This paper offers a precise, formal definition of an audio-to-score alignment.
While the concept of an alignment is intuitively grasped, this precision affords us new insight into the evaluation of audio-to-score alignment algorithms.
Motivated by these insights, we introduce new evaluation metrics for audio-to-score alignment. 
Using an alignment evaluation dataset derived from pairs of KernScores and MAESTRO performances,
we study the behavior of our new metrics and the standard metrics on several classical alignment algorithms.
\end{abstract}

\section{Introduction}\label{sec:intro}

Audio-to-score alignment is a fundamental problem in music information retrieval, with applications ranging from score following to music transcription. The concept of an alignment is typically given by an intuitive definition, for example:

\begin{enumerate}
\item ``Music alignment is the association of events in a score with points in the time axis of an audio signal.'' Orio and Schwarz (2001) \cite{orio2001alignment}.\label{itm:orio}

\item ``A procedure which, for a given position in one representation of a piece of music, determines the corresponding position within another representation.'' Ewert, M{\"u}ller, and Grosche (2009) \cite{ewert2009high}.\label{itm:ewert}

\item ``Match notes in a music performance signal (called an aligned signal) to those in a reference musical score or another performance signal (reference signal).'' Nakamura, Yoshii, and Katayose (2017) \cite{nakamura2017performance}.\label{itm:nakamura}
\end{enumerate}

We identify two problems with deferring to intuition for such a foundational concept. First, as we will discuss in Section \ref{sec:reldef}, these definitions are subtly inconsistent with each other. This is a potential source of confusion in how the community thinks and talks about alignments. Second, while the definition of an alignment is vague, the metrics used to evaluate alignments are quite precise. In the absence of an explicit definition of an alignment, these metrics become an implicit substitute for a definition. This is a problem if these metrics fail to capture the right intuitive concept. Furthermore, without a precise definition, it is difficult to critically analyze how well these metrics measure the concept of a good alignment.

This paper presents a precise, formal definition of an audio-to-score alignment: an increasing function that maps each location in the score to a time in the performance. Furthermore, we formalize what we mean by a \textit{good} audio-to-score alignment function, via metrics on the function space of all possible alignments. This allows us to design evaluation metrics that explicitly measure this goodness of alignment. Our treatment of alignments as functions is presented in Section~\ref{sec:func}, and metrics for evaluating the quality of an alignment function are presented in Section~\ref{sec:eval}.

The prevailing methodology for evaluating alignments artificially perturbs the source data, resulting in dis-aligned sequences that are then re-aligned by an alignment algorithm (see Section \ref{sec:reldata} for details). This is convenient, because we can easily quantify the accuracy of an alignment by measuring how well it inverts the dis-alignment procedure. But an artificial dis-alignment does not accurately model the expressive dis-alignment arising from a human performance of a score. This calls into question whether this evaluation procedure will reflect the reliability and accuracy of an alignment algorithm when it is applied to dis-alignments arising from expressive human performances.

In Section~\ref{sec:groundtruth} we introduce a new dataset that pairs scores with expressive human performances, together with ground-truth alignments between these pairs. This enables us to consider a realistic alignment task, with an evaluation methodology that eliminates the need for artificial perturbations. In Section~\ref{sec:results}, we evaluate several baseline alignment algorithms on this new dataset, and compare the metrics introduced in Section \ref{sec:eval} to the standard metrics. We find that our metrics are well-correlated with the standard metrics, and more suitable for evaluating the alignments discussed in Section \ref{sec:func}.

\section{Alignments as Functions}\label{sec:func}

Given a score of length $S$ and an expressive performance of this score of duration $T$, we define an audio-to-score alignment to be a function of the following form.
\begin{definition}\label{def:align}
A (temporal) audio-to-score alignment is a monotonic real function $\tau : [0,S) \to [0,T)$ that assigns each position $s \in [0,S)$ in the score to a time $t = \tau(s) \in [0,T)$ in the audio performance.
\end{definition}
This definition is expansive, intended to capture all functions that could meaningfully be interpreted as alignments. After some preliminary definitions in Section~\ref{sec:pianoroll}, we introduce the concept of the \textit{best}, ideal alignment between a score and a given performance in Section~\ref{sec:commute}. In Section~\ref{sec:reldef} we discuss another definition of an alignment, and draw a distinction between temporal alignments (Definition \ref{def:align}) and note-based alignments.

To build intuition for Definition \ref{def:align}, observe that an alignment function $\tau$ implicitly assigns notes in a score to time intervals in a performance. If a note spans the interval $[s_1,s_2)$ in the score, then it aligns to the interval $[\tau(s_1),\tau(s_2))$ in the performance. Monotonicity ensures that $\tau(s_1) \leq \tau(s_2)$. Mapping the timings of all the notes in the score to timings in the performance via an alignment function yields a performance-aligned score, a commonly studied object in the alignment literature that we define precisely in Section \ref{sec:pianoroll}. The relationship between a (piano-roll) score, a performance-aligned score, and an alignment is visualized in Figure~\ref{fig:alignment}.

\vspace{-2mm}
\subsection{Continuous Piano Rolls}\label{sec:pianoroll}

Let $N$ denote the number of distinct pitches that can be performed in a given genre of music. For solo piano music $N = 88$, the number of keys on a standard piano; most music in the Western semi-tone scale can be represented by $N=128$, the convention adopted by MIDI.

\begin{definition}\label{def:score}
A piano-roll of duration of $T$ is a discrete-valued function $\textbf{pianoroll} : [0,T) \to \{0,1\}^N$.
\end{definition}
If we discretize time $T$ into sub-divisions of length $dt$, then the matrix $M \in \{0,1\}^{N\times(T/dt)}$ defined by $M_{n,k} = \textbf{pianoroll}(kdt)_n$ is the more familiar, discrete-time version of a piano-roll.

We represent a score $\mathcal{S}$ as a piano-roll, denoted by $\textbf{score} : [0,S) \to \{0,1\}^N$, defined by the rule $\textbf{score}(s)_n = 1$ iff note $n$ is indicated by $\mathcal{S}$ at position $s \in [0,S]$. As a convention, we measure time in a score with units of beats; a score defined on the interval $[0,S)$ is $S$ beats long. A continuous piano-roll is a highly simplified representation of a musical score, but it captures the relevant features for audio-to-score alignment.

\begin{figure}
\centering
\includegraphics[width=\columnwidth]{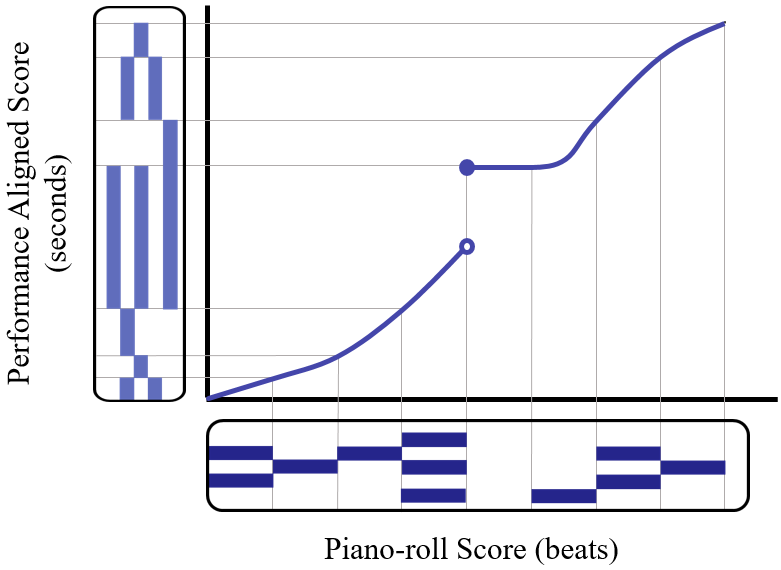}
\vspace{-6mm}
\caption{An alignment function $\tau$, that maps time in a score (measured in beats) to time in a performance (measured in seconds). A piano-roll \textbf{score} is mapped by the alignment to a performance-aligned score $\textbf{pscore}_\tau$. Time stops in the score as the performer extends the triad, resulting in a jump discontinuity in the alignment.}
\label{fig:alignment}
\vspace{-4mm}
\end{figure}

Given a piano-roll \textbf{score} and an alignment $\tau$ to a performance, we can construct a performance-aligned version of the score with expressive timings given by $\tau$.
\begin{definition}\label{def:alignedscore}
Given a piano-roll \textbf{score} and an alignment $\tau$, the associated performance-aligned score is given by the piano-roll $\textbf{pscore} : [0,T) \to \{0,1\}^N$ such that
\begin{equation*}
\textbf{pscore}_\tau(t) = \bigvee_{s\in\tau^{-1}(t)} \textbf{score}(s).
\end{equation*}
\end{definition}
The inverse image of $t$ in $\tau$, denoted by $\tau^{-1}(t)$, is the set of times in the score that $\tau$ aligns to time $t$ in the performance. Alignments $\tau$ need not be invertible, so the set $\tau^{-1}(t)$ may contain $0$, $1$, or many elements, as seen in Figure~\ref{fig:alignment}. Intuitively, the notes in a \textbf{pscore} at time $t$ are the union of notes denoted in the score at each time $s \in \tau^{-1}(t)$. We represent notes by indicators in a binary vector $\{0,1\}^N$, so this union can be represented as the bitwise logical disjunction of these vectors: $x_n \lor y_n = 1$ iff $x_n = 1$ or $y_n= 1$.

Another way to think about Definition \ref{def:alignedscore} is that we can elevate an alignment $\tau : [0,S) \to [0,T)$ to a function $\texttt{align}_\tau : \text{Piano Roll} \to \text{Piano Roll}$ defined by the rule $\texttt{align}_\tau(\textbf{score}) = \textbf{pscore}_\tau$. It is easy to conflate the real function $\tau$ with the lifted function $\texttt{align}_\tau$, or with the image $\textbf{pscore}_\tau$ of the score in this lifted alignment function; this may have contributed to the confusion about definitions that we alluded to in Section \ref{sec:intro}. We expand upon this discussion in Section \ref{sec:reldef}.

\vspace{-2mm}
\subsection{Ground-Truth Alignments}\label{sec:commute}

To evaluate the quality of a proposed alignment, we will want to compare to a ground-truth alignment: i.e. the ideal alignment function that correctly maps a score to a corresponding performance. Conceptually, this ideal alignment exists and is unique. To convince ourselves of this, we can imagine assigning the beginning of each beat in a score to a specific time in the performance. This assignment forms a crude, discretized alignment; by carefully interpolating this alignment to sub-beats and sub-sub-beats ad nauseam, we flesh out the full alignment function. Such thinking motivated the procedure of Raphael (2004) \cite{raphael2004hybrid}, who approximated ground-truth alignments from annotations given by a person, tapping along to the beat of a performance.

To give a precise definition of an ideal ground-truth alignment, we first introduce the concept of a symbolic performance. Suppose we are given an expressive performance $\mathcal{P}$ in a symbolic format. For example, $\mathcal{P}$ could be a symbolic sequence of notes recorded by a MIDI keyboard. In contrast to a score $\mathcal{S}$,  a performance has expressive timings, measured in units of seconds (see Figure~\ref{fig:alignment}). We will represent $\mathcal{P}$ as a piano-roll, denoted by $\textbf{symbp} : [0,T) \to \{0,1\}^N$, defined  by the rule $\textbf{symbp}(t)_n = 1$ iff note $n$ is being performed at time $t \in [0,T)$ in $\mathcal{P}$.

A performance-aligned score is a special case of a symbolic performance. Another example of a symbolic performance is the output of an ideal, frame-based music transcription system $\texttt{transcribe} : \text{Audio} \to \text{Piano Roll}$ which, at any point in an audio performance, indicates the set of notes being performed at that point in time. We consider the acquisition of symbolic performance transcripts in Section \ref{sec:groundtruth}. For now, we assume that we have access to an accurate transcription of a given audio performance. 

We can use symbolic performance transcripts to define ideal ground-truth alignments between scores and audio performances. Let $\texttt{perform} : \text{Score} \to \text{Audio}$ denote the action of a performer, who converts a score into audio through the act of performance. Borrowing the language of category theory \cite{riehl2017category}, we define an ideal ground-truth alignment to be an alignment function that makes the following diagram ``commute:''
\begin{center}
\vspace{-2mm}
\begin{tikzpicture}[baseline= (a).base]
\node[scale=1.0] (a) at (0,0){
\begin{tikzcd}[row sep=2.6em, column sep=6em, nodes={font={\fontsize{15pt}{15}\selectfont}}, labels={font={\fontsize{10pt}{10}\selectfont}}]
\substack{\text{Score}} \arrow[r,"\texttt{align}_{\tau^*}"]
                                     \arrow[dr,swap,"\texttt{perform}"]
& \substack{\substack{\text{Symbolic}\\\text{Performance}}} \\
& \substack{\text{Audio}}\arrow[u,swap,"\texttt{transcribe}"]
\end{tikzcd}
};
\end{tikzpicture}
\vspace{-2mm}
\end{center}
What we mean when we say that the diagram commutes is captured by the following definition.
\begin{definition}\label{def:gtalign}
An ideal ground-truth alignment between a score \textbf{score} and an audio performance $\texttt{perform}\hspace{.3mm}(\textbf{score})$ is an alignment $\tau^*$ such that
\[
\texttt{transcribe}(\texttt{perform}\hspace{.3mm}(\textbf{score})) = \texttt{align}_{\tau^*}(\textbf{score}).
\]
\end{definition}
In words: the ideal ground-truth alignment of a score to a given audio performance yields the same symbolic performance as a perfect transcript of the audio.

Unfortunately, the ideal ground-truth alignments defined in Definition \ref{def:gtalign} do not exist for most score-performance pairs, because human performances do not faithfully reproduce a musical score. In some cases, this is due to performance error: the insertion or deletion of a note that is not reflected in the score. Even for highly-accurate, professional performances, the deliberate extensions, asynchronies, and staccato articulations that comprise an expressive performance prevent an alignment that exactly satisfies equality in Definition \ref{def:gtalign}.

Mismatches between a performance-aligned score and a corresponding performance transcript are illustrated in Figure~\ref{fig:async}. Observe that, for the score and performance transcript illustrated in the figure, there is no alignment of the score that is equal to the given performance; among other problems, the missing note (marked completely in red) in the second-to-last beat of the performance cannot be avoided by any performance-aligned score constructed from a function given by Definition \ref{def:align}. We must therefore settle for an approximation to the equality in Definition \ref{def:gtalign}; we discuss how to obtain approximate ground-truth alignments in Section \ref{sec:groundtruth}.

\subsection{Related Definitions of an Alignment}\label{sec:reldef}

Definition \ref{def:align} formalizes the concept of a temporal alignment, described clearly by Ewert, M{\"u}ller, and Grosche \cite{ewert2009high} (quoted in Section \ref{sec:intro}, Item \ref{itm:ewert}). An alternate definition, that we will refer to as a note-based alignment, is eloquently stated by Nakamura, Yoshii, and Katayose \cite{nakamura2017performance} (quoted in Section \ref{sec:intro}, Item \ref{itm:nakamura}). Whereas Ewert et. al. view alignments as functions (i.e. ``procedures'') that map \textit{positions} in one sequence to another, Nakamura et. al. view alignments as functions (i.e. ``matchings'') that map \textit{notes} in one sequence to another. Both perspectives on alignment are common in the literature, although the definition of an alignment is often presented in more ambiguous terms (e.g. the definition quoted in Section \ref{sec:intro}, Item \ref{itm:orio}).

\begin{figure}
\centering
\includegraphics[width=\columnwidth]{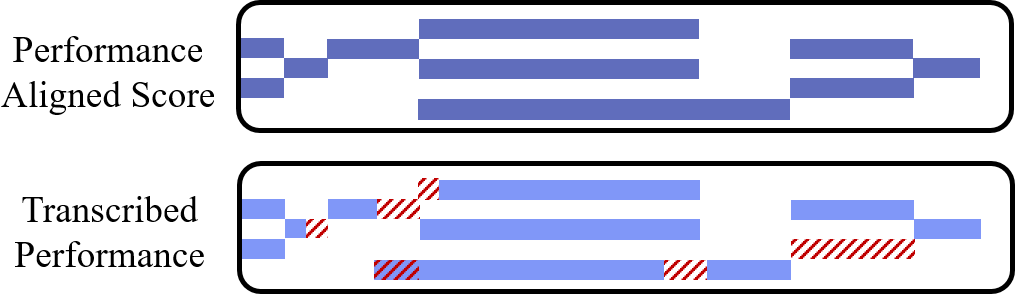}
\caption{Two symbolic performance piano-rolls. Top: a score is aligned to an audio performance to create a performance-aligned score. Bottom: the same audio performance is been transcribed to create a performance transcript. Asynchronies and errors in the pianist's performance prevent a perfect alignment (red dashes).}
\label{fig:async}
\vspace{-4mm}
\end{figure}

Imagine a performer strikes three notes on the fourth beat of a score asynchronously, as in Figure~\ref{fig:async}.
A temporal alignment necessarily maps the onsets of these three notes in the score to a single point of time in the performance. In contrast, a note-based alignment could map the onsets of these notes to distinct times in the performance. Note-based alignments have more flexibility to alter the structure of a score and satisfy  equality in Definition \ref{def:gtalign}. But flexibility comes at a cost. If the note-based alignment maps three notes in the triad to different times in the performance, then it cannot answer the following question: at what time in the performance is the fourth beat?

While we study temporal alignments in this paper, the note-based alignment problem is also important.
Audio-to-score alignment is commonly used as an intermediate tool to facilitate other MIR tasks: score-following, music transcription, beat tracking, and onset detection, among others. For tasks such as score following and and beat tracking, we want a temporal alignment that tell us what time a particular position in the score corresponds to in the performance. For tasks including frame-based music transcription and onset detection, we would prefer a note-based alignment that tells us where notes in the score occur in the performance.

\vspace{-2mm}
\section{Evaluating an Alignment}\label{sec:eval}

Having defined an alignment as a real-valued function, we propose to evaluate the quality of an alignment by measuring its distance (as a function) to a ground-truth alignment. In Section \ref{sec:defeval}, we introduce alignment metrics that measure the average error of an alignment function over the duration of a score. In Section \ref{sec:releval}, we compare these temporal metrics to note-based metrics commonly reported in the literature.

\vspace{-2mm}
\subsection{Alignment Metrics}\label{sec:defeval}
\vspace{-1mm}

\begin{figure}
\centering
\includegraphics[width=\columnwidth]{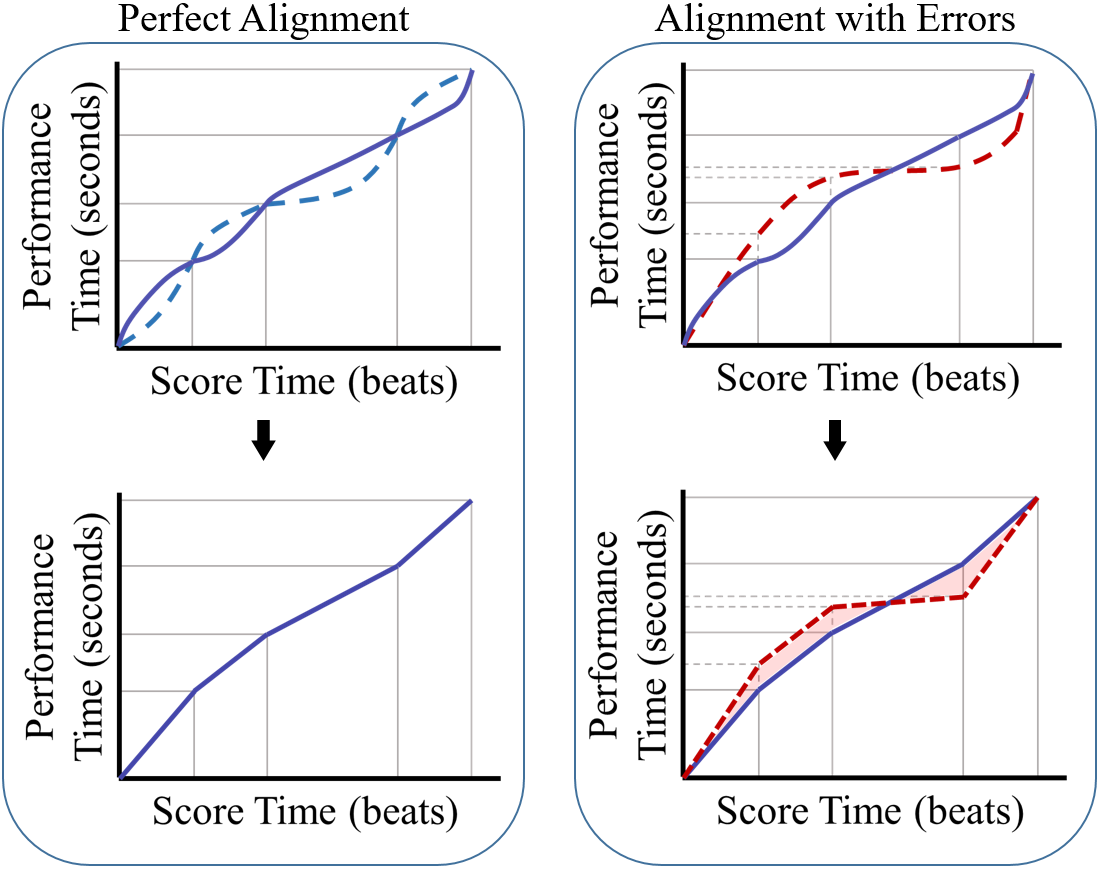}
\caption{A possible ground-truth alignment (solid) compared to two candidate alignments (dashed), shown before (top) and after (bottom) linearization. The first candidate alignment (left) is perfect; its canonical linear representative exactly matches that of the ground-truth. The second candidate alignment (right) maps score changepoints erroneously; its error is given by the shaded region.}
\label{fig:twoLinearizations}
\vspace{-3mm}
\end{figure}

We first observe that there is an inherent alignment ambiguity between changepoints in a score. During these intervals, time evolves even though the set of notes being played does not change. The path the alignment takes between these two score events is therefore underspecified. This is demonstrated in the left panel of Figure \ref{fig:twoLinearizations}, where the ground-truth and candidate alignments satisfy $\tau(s_i) = \tau^*(s_i)$ at each changepoint $s_i$, but evolve differently between changepoints. We consider two alignments $\tau_1,\tau_2$ \textit{equivalent} if $\tau_1(s_i) = \tau_2(s_i)$ for every changepoint $s_i$ in the score. To compare two alignments, we will compare canonical representatives of their equivalence classes. As a canonical representative, we pick the alignment that linearly interpolates between changepoints; we write $\tilde\tau$ to denote this linearization of an alignment function $\tau$. Linear interpolation is a natural choice because it represents time evolving at a constant pace between changepoints.

We define the temporal error of an alignment $\tau$, given ground-truth $\tau^*$, by the $L^1$ distance between $\tau$ and $\tau^*$.
\begin{definition}\label{def:mad}
Temporal average error ($\mad$) between alignments $\tau$ and $\tau^*$ is given by
\vspace{-1mm}
\[
\mad(\tau, \tau^*) \equiv \frac{1}{S}\int_0^S |\tilde\tau(s) - \tilde\tau^*(s) | ds.
\]
\end{definition}
At any point $s$ in the score, the quantity $|\tilde\tau(s) - \tilde\tau^*(s)|$ measures how far the candidate alignment's position in the performance, $\tilde\tau(s)$, differs from the ground-truth time in the performance, $\tilde\tau^*(s)$. $\mad$ computes the average of these errors over score time.

The $\mad$ metric alone does not distinguish between different distributions of the alignment error. For example, one alignment may lag by 10ms for the duration of the performance, while another may have one measure with large error in an otherwise perfect alignment. To distinguish between these types of errors, we also consider the standard deviation of errors over the entire score.
\begin{definition}\label{def:rmse}
Temporal standard deviation ($\rmse$) between alignments $\tau$ and $\tau^*$ is given by
\[
\rmse(\tau, \tau^*) \equiv \sqrt{\frac{1}{S}\int_{0}^S \left(\tilde\tau(s)) - \tilde\tau^*(s) \right)^2ds}. 
\]
\end{definition}
Because the alignments are linearized, both $\mad$ and $\rmse$ can be calculated in closed form. To avoid extraneous alignment errors involved in guessing the offset time of the final notes of a performance, we define the end of the score $S$ to be the time of the last \textit{onset} in the score.

Finally, we note that the alignment paths produced by common alignment algorithms, including dynamic time warping, are not necessarily functions. In particular, they may not be invertible; one score event may map to multiple events or even an interval in the performance. In these cases, we choose to treat the alignments as right-continuous functions (as illustrated in Figure \ref{fig:alignment}).

\vspace{-2mm}
\subsection{Related Alignment Metrics}\label{sec:releval}
\vspace{-1mm}

The metrics introduced in Section \ref{sec:defeval} measure temporal differences in the performance with respect to \textit{positions} in the score. In contrast, audio-to-score alignment are usually evaluated based on temporal differences in the performance with respect to \textit{notes} in the score. These note-based metrics are clearly described in Cont et. al. \cite{cont2007evaluation}, although similar metrics appeared in earlier audio-to-score alignment papers \cite{meron2001automatic,orio2001alignment,shalev2004learning}. The most commonly reported metric is the mean absolute difference between aligned note onsets and ground-truth onsets\cite{meron2001automatic,orio2001alignment,shalev2004learning,keshet2007large,ewert2008refinement,ewert2009high,devaney2009handling,devaney2014estimating,niedermayer2009improving,niedermayer2010multi,lajugie2016weakly,kwon2017audio,arzt2018audio}. This metric is a note-based analog to Definition \ref{def:mad}; given a list of $N$ onsets $\textbf{s}$ in a score corresponding to onsets $\textbf{p}$ in a performance, we can compute
\vspace{-1mm}
\begin{equation}
\notemad(\textbf{s},\textbf{p}) \equiv \frac{1}{N}\sum_{i=1}^N |\textbf{s}_i - \textbf{p}_i|.\label{eqn:notemad}
\end{equation}
Some works additionally report mean deviation of note-offsets, and standard deviations of onset and offset times. The standard deviation of onset times is a note-based analog to Definition \ref{def:rmse}:
\vspace{-3mm}
\begin{equation}
\notermse(\textbf{s},\textbf{p}) \equiv \sqrt{\frac{1}{N}\sum_{i=1}^N \left(\textbf{s}_i - \textbf{p}_i\right)^2}.\label{eqn:notermse}
\end{equation}
Another commonly reported metric is the fraction of onsets that exceed some threshold deviance from ground-truth, sometimes called the ``onset recognition rate'' \cite{cont2007evaluation,joder2011conditional,joder2013learning,joder2013off,carabias2015audio}.

In light of the discussion in Section \ref{sec:reldef}, there is an evident mismatch between the concept of a temporal alignment and these note-based evaluation metrics. Indeed, we argue that these note-based metrics have pushed researchers to overly focus on the note-based alignment problem: the flexibility to adjust the onset times of individual notes allows note-based alignment algorithms to easily outperform temporal alignment algorithms when we measure them using note-based metrics. Even Ewert's work \cite{ewert2008refinement,ewert2009high}, which clearly states its purpose to construct temporal alignments, evaluates using the note-based metrics.

\section{Evaluation Methodology} \label{sec:groundtruth}

To evaluate alignment algorithms using the metrics introduced in Section \ref{sec:eval} requires a dataset consisting of:
\begin{enumerate}
\item Musical scores in a symbolic digital format.\label{itm:score}
\item Relatively faithful performances of these scores.\label{item:perf}
\item Ground-truth alignment annotations.\label{item:gt}
\end{enumerate}
In Section \ref{sec:dataset} we introduce a dataset that satisfies these requirements, derived from the public KernScores \cite{sapp2005online} and MAESTRO \cite{hawthorne2018enabling} datasets. The chief technical difficulty in constructing this dataset involves constructing the ground-truth alignment annotations; in Section \ref{sec:optgt} we propose methodology for approximating the ideal ground-truth alignments between KernScores and MAESTRO performances; the key insight is that, while constructing a temporal alignment between a score and an \textit{acoustic} performance is difficult, aligning a score to a \textit{symbolic} performance transcript is much easier.\footnote{For this reason, works that study score-to-MIDI alignment problem focus on note alignments, which present a challenge even in the symbolic setting \cite{nakamura2017performance}.} In Section \ref{sec:reldata} we review other commonly-used datasets for evaluating alignments.

\subsection{A Dataset of Ground-Truth Alignments}\label{sec:dataset}

For the results presented in Section~\ref{sec:results}, we constructed a dataset of ground-truth alignments by cross-referencing a subset of the KernScores collection of musical scores with a subset of the MAESTRO v2.0.0 dataset of piano performances and transcripts. The KernScores dataset contains scores for the complete collection of Preludes and Fugues from Bach's Well-Tempered Clavier, which can be cross-referenced to performances in the MAESTRO dataset. This subset consists of 193 performances, with a total of 440 minutes of audio. One notable property of this dataset is that the performers have a range of skill, and some performances contain substantial errors. This creates an interesting challenge for alignment algorithms. A script to reconstitute this dataset from copies of KernScores and MAESTRO is provided in the repository (see Section \ref{sec:results}).

\subsection{Approximating the Ground-Truth Alignment}\label{sec:optgt}

In order to quantify an approximation to the equality in Definition \ref{def:gtalign}, we must define a metric on the space of symbolic performances. We propose using the $L^1$ distance.
\begin{definition}\label{def:metric}
Given two symbolic performances $\textbf{sp}_1$ and $\textbf{sp}_2$ of duration $T$, the $L^1$ distance between them is
\[
d_1(\textbf{sp}_1,\textbf{sp}_2) = \frac{1}{T}\int_0^T \|\textbf{sp}_1(t) - \textbf{sp}_2(t)\|_1\,dt.
\]
\end{definition}
Intuitively, $\|\textbf{sp}_1(t) - \textbf{sp}_2(t)\|_1$ counts the number of differences between $\textbf{sp}_1$ and $\textbf{sp}_2$ at an instant in time $t$, and the integral computes the cumulative average difference. Finding $\tau$ that minimizes $d_1(\textbf{pscore}_\tau, \textbf{transcript})$, approximating the equality in Definition~\ref{def:gtalign}, can be achieved by classical dynamic time warping in $O(ST)$ time and space \cite{sakoe1978dynamic}.

We define the best approximation to the ideal alignment given by Definition~\ref{def:gtalign} as the solution to the following regularized optimization problem:
\begin{equation}\label{eqn:optgt}
\begin{array}{ll@{}ll}
\underset{\tau : [0,S] \to [0,T]}{\text{minimize}} &d_1(\textbf{pscore}_\tau, \textbf{transcript}) + \lambda R(\tau), \\
\text{subject to}& \tau(s_1) \leq \tau(s_2)  \text{ if $s_1 < s_2$}. & 
\end{array}
\end{equation}
The constraints force $\tau$ to satisfy the definition of an alignment (Definition \ref{def:align}). The term $d_1(\textbf{pscore}_\tau, \textbf{transcript})$ forces $\tau$ to approximate the equality in Definition~\ref{def:gtalign}, and can be interpreted as a measure of compatibility of the alignment $\tau$ with the performance. The term $\lambda R(\tau)$ regularizes the optimization towards a uniform tempo, and can be interpreted as a measure of compatibility of $\tau$ with the score. For the experiments in Section~\ref{sec:results} we set $\lambda = 0.1$. 

Let $\rho(\tau)$ be the mean inverse-tempo of $\tau$ (measured in seconds per beat). We use the variance of the inverse-tempo, $R(\tau)$, to regularize $\tau$ towards its mean tempo:
\begin{equation}
R(\tau) \equiv \frac{1}{S} \int_0^S \left(\frac{d\tau(s)}{ds} - \rho(\tau)\right)^2 \,ds.
\end{equation}
By the fundamental theorem of calculus, we can write the mean inverse-tempo of an alignment as
\begin{equation}
\rho(\tau) \equiv \frac{1}{S} \int_0^S \frac{d\tau(s)}{ds}\,ds = \frac{\tau(S) - \tau(0)}{S} = \frac{T}{S}.
\end{equation}
Crucially, the mean inverse-tempo is independent of the path $\tau$, which enables us to write a dynamic program--analogous to dynamic time warping--to compute the minimization problem Equation~\ref{eqn:optgt} in $O(ST^2)$ time using $O(ST)$ space.

Computing the note-based metrics discussed in Section \ref{sec:releval} requires a correspondence between notes in the score and notes in the performance. We construct this correspondence from the ground-truth alignment using a heuristic. For each note in the score, we map its onset time to a time $t$ in the performance via the ground-truth alignment. The closest note in the performance transcript with the same pitch is matched to the note in the score. If no note of the same pitch is found within $\pm 100$ms of $t$, the note in the score is considered unmatched and excluded from note-based calculations; using the $100$ms threshold, we achieve a $96.7\%$ correspondence between notes on this dataset.

\subsection{Related Alignment Datasets}\label{sec:reldata}

A common way to evaluate alignments is to construct a synthetic dataset using a synthesizer\cite{meron2001automatic,orio2001alignment,hu2003polyphonic,maezawa2015bayesian,lajugie2016weakly}.
While this method is convenient, results on synthesized performances could mislead us if we want to understand how an algorithm will behave on human performances. Furthermore, a score cannot be directly synthesized in this approach: most synthesizers produce inexpressive, constant-tempo performances, and aligning to such performances is trivial. The typical solution is to perturb note onsets and offsets in the score before synthesizing \cite{ewert2012towards,raffel2016optimizing}, but the artificial nature of these perturbations could also make the results misleading.

Another way to constructing a dataset uses performance transcripts captured during an acoustic piano performance using a sensor array that records key and pedal presses \cite{soulez2003improving,shalev2004learning,keshet2007large,niedermayer2009improving,niedermayer2010multi,joder2010improved,joder2011conditional,joder2013learning,joder2013off,kwon2017audio}. Collections of these performances and transcripts have been made widely available by the MAPS \cite{emiya2009multipitch} and MAESTRO datasets. These datasets do not include corresponding scores and, like the process for synthesized datasets, common practice is to randomly adjust the timings of the performance transcripts to create a ``score'' for input to the alignment algorithm. These adjusted transcripts are unlikely to look anything like actual scores: this was our motivation to identify a subset of MAESTRO corresponding to genuine scores in Section~\ref{sec:dataset}. A limitation of using sensor-derived performance transcripts to construct a dataset (including our dataset) is that it restricts evaluation to piano music.

Finally, a labor-intensive approach dataset construction involves gathering scores, corresponding performances, and manually annotating the ground-truth alignment. This approach was taken for the Bach10 dataset \cite{duan2010multiple}. Other datasets have been constructed this way, and are used for the MIREX Real-time Audio to Score Alignment task \cite{miron2014audio,arzt2018audio}, but these datasets do not appear to be public; this is understandable, as the effort required to create this data means the datasets are most valuable as privately-held test sets. A limitation of manual annotation is that the cost to label data is so high that it is unlikely large datasets of this sort will ever become available; the Bach10 dataset consists of 10 short recordings of Bach Chorales, totaling 5.5 minutes of music. Larger datasets are desirable, in particular for alignment algorithms based on machine learning.

\vspace{-2mm}
\section{Conclusion}\label{sec:results}

To understand the relationship between the temporal metrics introduced in Section \ref{sec:defeval} and the note-based metrics discussed in Section \ref{sec:releval}, we compute both sets of metrics for alignments of the dataset introduced in Section \ref{sec:dataset}. Because we focus on the temporal alignment task, the relevant alignment algorithms are variants of dynamic time warping. We consider three classic variants of DTW, using features derived from a synthesizer \cite{turetsky2003ground}:
\begin{enumerate}
\setlength\itemsep{.3em}
\item Alignment with (log-)spectrogram features.
\item Alignment with (log-)chromagram features \cite{hu2003polyphonic}.
\item Alignment with constant-Q transform features \cite{raffel2016optimizing}.
\end{enumerate}
We compute features from a synthesized performance of the score, created using PrettyMidi's FluidSynth interface \cite{raffel2014intuitive}, with a hop-size of $512$ samples ($\approx 12$ms). We use librosa's \cite{mcfee2015librosa} implementation of dynamic time warping to compute an alignment between featurizations. For the Constant-Q featurization, we use Raffel and Ellis's hyper-parameter settings \cite{raffel2016optimizing}, however we do not apply their gully and penalty constraints as we found that these degrade results for performances that take a very different tempo than the corresponding score.

\begin{table}[!t]
\begin{center}
  \setlength\tabcolsep{4pt}
  \begin{tabular}{ | l || c | c | c | }
    \hline
     & Spectra & Chroma & CQT \\ \hline\hline
     $\mad$ (Definition \ref{def:mad}) & 37 & 35 & 33   \\ \hline
     $\notemad$ (Equation \ref{eqn:notemad}) & 25 & 26 & 23 \\ \hline\hline
     $\rmse$ (Definition \ref{def:rmse}) & 114 & 76 & 97 \\ \hline
     $\notermse$ (Equation \ref{eqn:notermse}) & 92 & 64 & 66  \\ \hline
  \end{tabular}
\end{center}
\vspace{-5mm}
  \caption{The average value of each metric across all performances in the dataset. Values are reported in milliseconds of performance time (lower is better). Spectrogram results exclude 13 outliers with $\mad$ $>$ 300ms.} 
  \label{tab:results}
  \vspace{-2mm}
\end{table}

\begin{table}[!t]
\begin{center}
  \setlength\tabcolsep{4pt}
  \begin{tabular}{ | l || c | c | c | }
    \hline
     & Spectra & Chroma & CQT \\ \hline\hline
     $\mad \text{vs.} \notemad$ & .94 & .93 & .87   \\ \hline
     $\rmse \text{vs.} \notermse$ & .85 & .83 & .85 \\ \hline
  \end{tabular}
\end{center}
\vspace{-5mm}
  \caption{Correlation coefficients between temporal alignment metrics and the analogous note-based alignment metrics. There is substantial agreement between these two sets of metrics about whether a performance is well-aligned.} 
  \label{tab:correlation}
  \vspace{-5mm}
\end{table}

We compare $\mad$ and $\rmse$ to the analogous note-based metrics $\notemad$ and $\notermse$. To calculate the note-based metrics, we use the correspondence between notes in a score and notes in a performance established in Section \ref{sec:optgt}. Results are summarized in Table~\ref{tab:results}. We find that the chroma and constant-Q alignments have comparable performance. This is consistent with previously reported results on synthetic data \cite{raffel2016optimizing}. Spectrogram alignments were prone to catastrophic failure; we removed 13 outlier spectrogram alignments from our evaluation, for which $\mad$ $>$ 300ms. Even on the remaining 182 performances, the spectrogram alignments are substantially worse than chroma or constant-Q.

Table \ref{tab:correlation} shows that the correlation between the temporal and note-based metrics is high. This assuages concerns that note-based metrics could be overly sensitive to regions of a score with a high density of note onsets: rapid processions of short-duration notes, or regions of high polyphony. While they correlate well, we recommend the new time-based metrics for the temporal alignment task for two reasons. First, temporal metrics eliminate the need for the ad-hoc thresholding heuristic discussed in Section \ref{sec:optgt}. Second,  the temporal metrics narrowly target the temporal alignment task: this removes the temptation to make misleading comparisons with results for the note alignment task, or to shift focus to the note alignment task in order to boost performance on the note-based alignment metrics.

The dataset introduced in this paper, along with code for evaluating alignments, is available on GitHub.\footnote{\url{https://github.com/jthickstun/alignment-eval}} We also provide the code for generating the dataset, including an implementation of the tempo regularization algorithm discussed in Section \ref{sec:optgt}. Tempo regularization can be seen as a principled alternative to gully or penalty methods \cite{raffel2016optimizing} and could be of general interest as a method for regularizing alignments. Finally, while we constructed this dataset for the purpose of evaluating alignments, it could be repurposed for other tasks that require high-quality alignments between performances and scores.

\clearpage

\bibliography{references}

\appendix
\onecolumn

\section{Visualizing Alignments}

\begin{figure}[h!]
\centering
\includegraphics[width=.92\columnwidth]{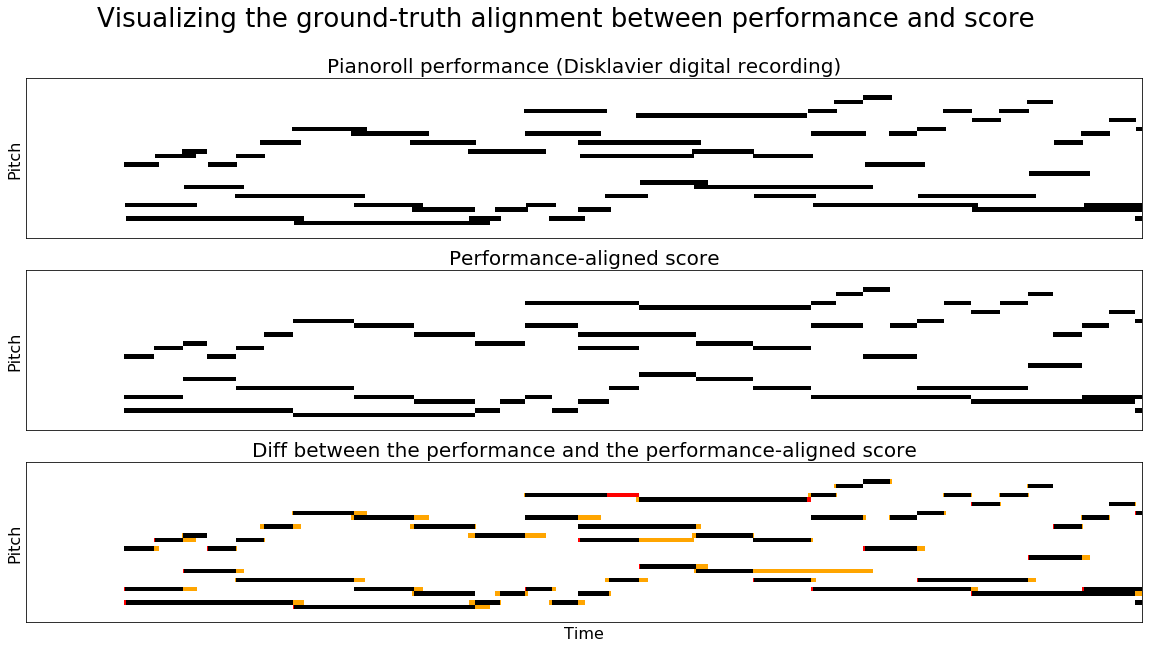}
\caption{To understand the behavior of the ground-truth alignments, we can visually compare the piano-roll performance (top) captured by the Yamaha Disklavier to the performance-aligned score created by warping the score according to the ground-truth alignment (middle). In the comparison plot (bottom) we use red to identify notes that are indicated by the performance-aligned score but not performed and yellow to identify notes that are performed but not indicated by the performance-aligned score. This example visualizes the beginning of a performance of the Bach's Prelude and Fugue in G-sharp minor (BWV 863).}
\label{fig:visualgt}
\end{figure}

\begin{figure}[h!]
\centering
\includegraphics[width=.92\columnwidth]{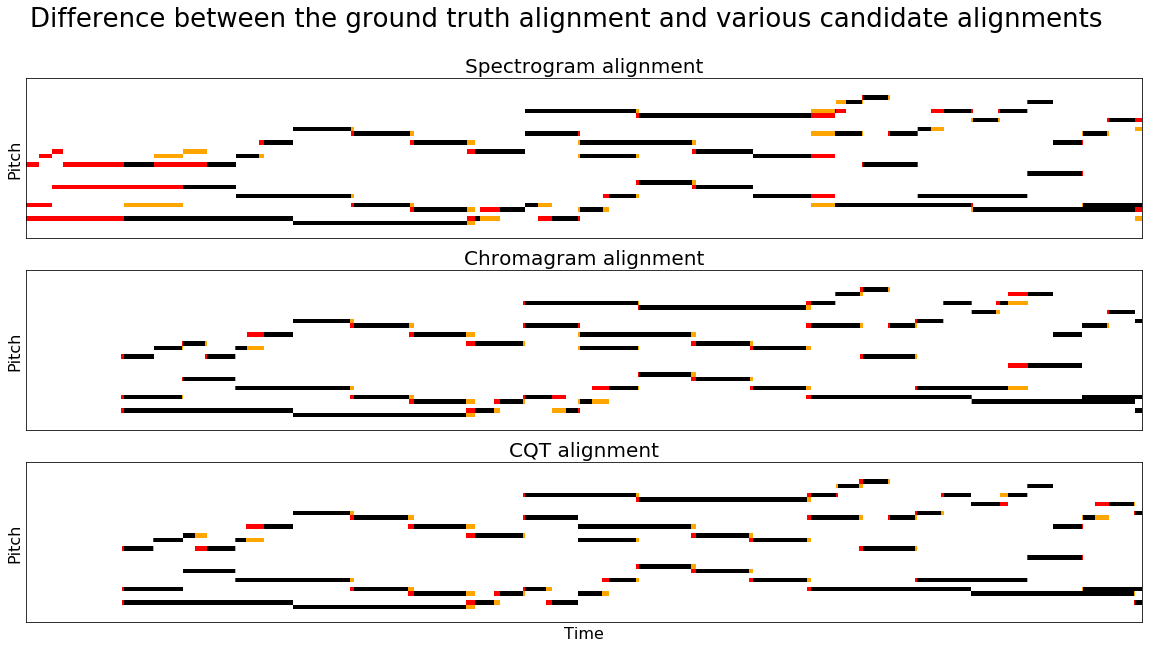}
\caption{We can also use these visualizations to compare the results of an candidate alignment algorithm to the ground-truth alignment. In each case, red is used to identify notes that are indicated by the candidate alignment algorithm, but not by the ground-truth alignment, and yellow is used to identify notes that are indicated by the ground-truth alignment, but not by the candidate alignment. This example visualizes the beginning of a performance of the Bach's Prelude and Fugue in G-sharp minor (BWV 863).}
\label{fig:visualalign}
\end{figure}

\section{A Visual Comparison of Time and Note Based Metrics}

\begin{figure}[h!]
\centering
\includegraphics[width=\columnwidth]{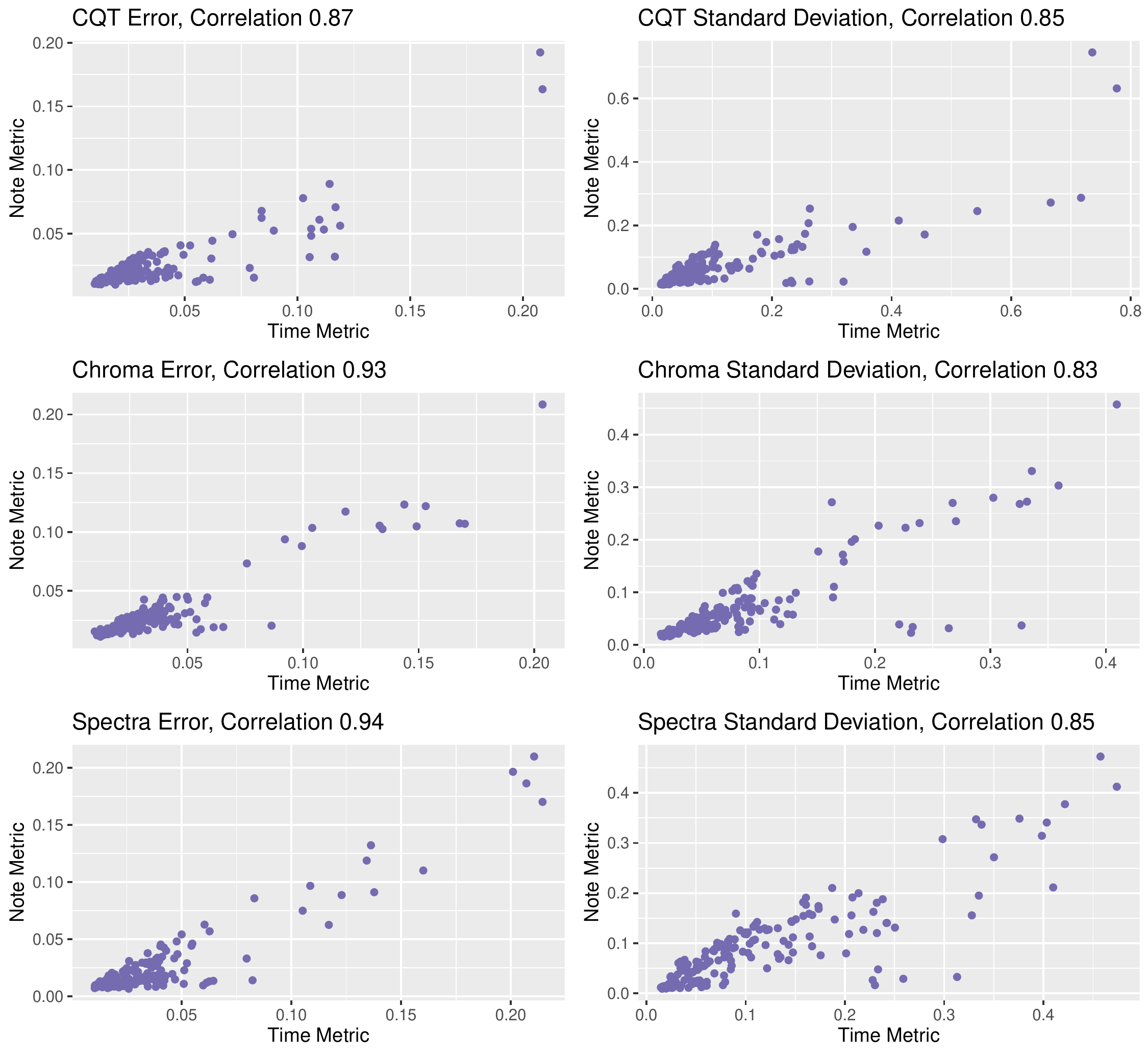}
\caption{A visual illustration of correlation between the new temporal metrics with the old note-based metrics. Each point is one of the 193 performances in the dataset. For spectrogram results, 13 outliers with $\mad$ $>$ 300ms are omitted from the plots.}
\label{fig:correlations}
\end{figure}

\end{document}